\documentclass[12pt]{article}
\usepackage{sao1}
\usepackage{psfig}

\begin{document}

 \begin{center}
{\LARGE \bf $H\alpha$ Survey of the Local Volume: Isolated Southern Galaxies}

\bigskip

{\large S. S. Kaisin$^1$,
A. V. Kasparova$^2$,
A. Yu. Knyazev$^{1, 3, 4}$,
and I. D. Karachentsev$^1$}

\bigskip

$^1$Special Astrophysical Observatory, Russian Academy of Sciences, Nizhnii Arkhyz, Karachai-Cherkessian Republic, 369167 Russia \\
$^2$ Sternberg Astronomical Institute, Universitetskii pr. 13, Moscow, 119992 Russia                                               \\
$^3$ European Southern Observatory, Garching, Germany                                                                                \\
$^4$ South African Astronomical Observatory, Cape Town, 7925 South Africa                                                              \\
\end{center}
Received June 13, 2006.

\bigskip

{\large \bf Abstract}

We present our $H_\alpha$ observations of 11 isolated southern galaxies:
 SDIG, PGC~51659, E~222-010, E~272-025, E~137-018,
IC 4662, Sag DIG, IC 5052, IC 5152, UGCA 438, and E~149-003, with
distances from 1 to 7 Mpc. We have determined the total $H_\alpha$ fluxes
from these galaxies. The star formation rates in these galaxies range from
$10^{-1}$ (IC~4662) to $10^{-4}M_{\odot}$ yr$^{-1}$ (SDIG) and the gas
depletion time at the observed star formation rates lies within
the range from 1/6 to 24 Hubble times $H_0^{-1}$ .

{\em Keywords: galaxies, groups of galaxies,
interstellar gas in galaxies, galaxy evolution.}

\section{INTRODUCTION}
Our views of the star formation in galaxies still remain
fragmentary. Until recently, the studies have been focused mainly on the
brightest objects. Dwarf systems, particularly those of low surface
brightness, remained out of sight, primarily because they were
difficult to observe. At the same time, a complete picture
cannot be obtained if dwarf galaxies are ignored. Being the elements
from which large systems were formed through mergers, they play a key
role in the formation and evolution of galaxies. Dwarf galaxies are
also interesting in that they are the most commonly encountered type
of galaxies in the Universe. Clearly, these objects are easiest to
study in the Local Volume.

The most complete sample of the Local Volume
is presented in the Catalog of Nearby Galaxies (Karachentsev et al. 2004).
This catalog contains 450 galaxies whose distances do not exceed 10 Mpc.
In recent years, about 30 more hitherto unknown nearby dwarf systems
have been discovered. An enormous work has been performed to refine
the distances of these galaxies based mainly on the luminosities of their
red giant branch stars.

The fact that there are no spiral density waves
in dwarf galaxies is important for studying the star formation.
Nevertheless, irregular galaxies have star
formation rates per unit luminosity that are approximately the same
as those of spiral galaxies (Hunter and Gallagher 1986). This paper
is a continuation of our previous works (Karachentsev et al. 2005;
Kaisin and Karachentsev 2006) on the $H\alpha$ survey of Local Volume
galaxies.

\section{OBSERVATIONS AND DATA REDUCTION}
We obtained CCD images in
the $H\alpha$ line and in the neighboring continuum for 11 galaxies of
the Local Volume. The observations were performed at the 2.2m ESO
telescope from July through August 2004. The images were taken with
the Wide Field Imager, which consists of eight CCD arrays with a
total field of view of $33'$x$34'$ and a resolution of
0.238$^{\prime\prime}$
per pixel.

The data were obtained with two narrow-band filters.
Of these, $H\alpha$/7 (ESO~856) was centered
at $\lambda = 6588.27$\AA\, and had a width
of $\Delta\lambda = 74.31$\AA.\,
The continuum \,images\, were\, taken\, with the ESO~858
filter
cen\-te\-red at $6655.61$\AA\,
 with a width of 120.78\AA.\,
The typical exposures were
800 and 300 s in the $H\alpha$ line and the continuum,
respectively.

The data
were reduced using the standard MIDAS package. Each image was reduced as
follows. First, we subtracted a bias from all images and then divided
them by fields, which were taken on the same observing
night as the galaxy in almost all cases. After automatic cosmic-ray hit
removal, the images
were corrected for scale distortions. To this end,
the field in the immediate neighborhood of each galaxy
was fitted by a second-degree polynomial.

We obtained
the final $H\alpha$ images by reducing the $H\alpha$ and continuum
images to the same scale and by applying corrections for airmass and
exposure time. To calibrate the fluxes, we used spectrophotometric
standard stars observed on the same nights as the objects. Some of our
images exhibit ring-shaped reflexes from bright stars. We removed
these artifacts when determining the $H\alpha$ fluxes from the galaxies.

\section{RESULTS}
 Table 1 present the main characteristics of the 11 galaxies that we
observed. Its columns contain the following: galaxy names (numbers) in known
catalogs, equatorial coordinates for the epoch 2000.0, standard angular
diameters in arcmin and apparent axial ratios, morphological types in de
Vaucouleur's digital code, apparent $B$ magnitudes, and distance to
the galaxies in Mpc. References to the original data are given in the
catalog by Karachentsev et al. (2004).

\begin{table}[hbtp]
\caption{}
\begin{center}
\begin{tabular}{lccccccc}

\hline
\hline
Object & RA & DEG & $a$ & b/a & Type & $B_t$ & D \\
  & h m s  & o ' " & $arcmin$ & & & $mag$ & Mpc \\
\hline
E349-031, SDIG  & 00 08 13.3 & -34 34 42 & 1.1 & 0.82 & 10 & 15.48 & 3.21 \\
PGC51659        & 14 28 03.7 & -46 18 06 & 2.4 & 0.38 & 10 & 16.50 & 3.58 \\
E222-010        & 14 35 03.0 & -49 25 18 & 0.9 & 0.44 & 10 & 16.33 & 5.8 \\
E272-025        & 14 43 25.5 & -44 42 19 & 3.0 & 0.50 &  8 & 14.77 & 5.9 \\
E137-018        & 16 20 59.3 & -60 29 15 & 3.2 & 0.34 & 9  & 12.20 & 6.34 \\
IC4662, E102-14 & 17 47 06.3 & -64 38 25 & 2.8 & 0.57 & 9  & 11.74 & 2.44 \\
SagDIG, E594-4  & 19 29 59.0 & -17 40 41 & 2.9 & 0.72 & 10 & 14.12 & 1.04 \\
E074-15, I5052  & 20 52 06.2 & -69 12 14 & 5.9 & 0.14 & 7  & 11.68 & 6.03 \\
E237-027, I5152 & 22 02 41.9 & -51 17 43 & 5.2 & 0.62 & 10 & 11.06 & 2.07 \\
UGCA438         & 23 26 27.5 & -32 23 26 & 1.5 & 0.80 & 10 & 13.86 & 2.23 \\
E149-003        & 23 52 02.8 & -52 34 39 & 2.2 & 0.18 & 9  & 15.0  & 6.4 \\
\hline
\hline

\end{tabular}
\end{center}
\end{table}

\begin{table}
\caption{}
\begin{center}
\begin{tabular}{l r r r r r r r r r r}

\hline
\hline
Object & $M_B$ & TI & $M_{HI}/L$ & $A_{H\alpha}$ & $lg F_{c}$ & lg SFR & $p_*$ & $f_*$ \\
(1)    & (2) & (3)&    (4)     &   (5)         &   (6)      &   (7)  & (8) & (9) \\
\hline
SDIG     & -12.10 & -0.8 & 0.55 & 0.03 & -14.02 & -3.89 & -0.75 &  1.35\\
PGC51659 & -11.83 &  0.0 & 6.6  & 0.30 & -13.68 & -3.47 & -0.22 &  1.04\\
E222-010 & -13.60 & -1.4 & 0.84 & 0.60 & -12.88 & -2.25 &  0.29 & -0.36\\
E272-025 & -14.77 & -1.5 & 0.21 & 0.37 & -12.80 & -2.15 & -0.08 & -0.60\\
E137-018 & -18.12 & -1.8 & 0.13 & 0.57 & -11.87 & -1.17 & -0.44 & -0.44\\
IC 4662  & -15.56 & -0.7 & 0.72 & 0.16 & -10.91 & -1.04 &  0.72 & -0.82\\
SagDIG   & -11.49 & -0.3 & 1.1  & 0.28 & -13.09 & -3.96 & -0.58 &  0.62\\
IC 5052  & -18.13 & -2.2 & 0.35 & 0.12 & -11.71 & -1.05 & -0.32 & -0.14\\
IC 5152  & -15.67 & -1.1 & 0.36 & 0.06 & -11.80 & -2.06 & -0.35 & -0.09\\
UGCA438  & -12.94 & -0.7 & 0.81 & 0.03 & -13.51 & -3.72 & -0.92 &  0.83\\
E149-003 & -14.09 & -1.7 & 0.89 & 0.03 & -12.90 & -2.19 &  0.15 & -0.20\\

\hline
\hline

\end{tabular}
\end{center}
\end{table}

Figure 1 shows $H\alpha$ plus
continuum (left) and $H\alpha$ minus continuum (right) images of the
galaxies from our sample. All images are 800$\times$800 pixels in
size, except for the two most extended galaxies ESO~074-15 and
ESO~237-27 with image sizes of 1300$\times$1300 pixels. Table 2
gives some of the calculated integrated parameters for the observed
galaxies: (1) galaxy
names; (2) their absolute magnitudes corrected for the Galactic
extinction as prescribed by Schlegel et al. (1998) and for the
internal extinction as described in the Catalog of Nearby Galaxies;
(3) the ``tidal index'' $TI$ introduced in the Catalog
of Nearby Galaxies to characterize the galaxy's surroundings
(negative $TI$ correspond to field galaxies for which the
tidal effect of neighbors is negligible); (4) the ratio of
the HI mass to the blue luminosity in solar units from the Catalog
of Nearby Galaxies corrected for new distance estimates; (5)
Galactic extinction $H\alpha/7$ line, $A(H\alpha)=2.32\cdot E(B-V)$,
where $E(B-V)$  is the color excess from Schlegel et al. (1998);
(6) extinction-corrected logarithm of the $H\alpha$ flux from the
galaxy, $F_c$ , in erg cm$^{-2}$s$^{-1}$; and (7) star formation
rate in the galaxy on a logarithmic scale defined as
$SFR(M_{\odot}$/year) = $1.27\cdot 10^9 F_c(H\alpha)\cdot D^2$,
(Gallagher et al.
1984), where the distance to the galaxy is in Mpc. The last two
columns give the dimensionless parameters $p_*$ and $f_*$
(see the next section) that which characterize the global star
formation activity of the galaxy in the past and in the future,
respectively. Note some of the individual properties of the observed
galaxies.

{\bf E~349-031 = SDIG (Sculptor Dwarf Irregular Galaxy).}
This low surface brightness galaxy is located in the region of a sparse
group (filament) in Sculptor. Karachentsev et al. (2006a) recently
determined its distance from the luminosities of red giant branch (RGB)
stars. The nearest significant
neighbors of SDIG are NGC 7793 and NGC 253, the ``crossing
time'' with which exceeds the Hubble time $T_0 = H_0^{-1}$.
Miller (1996) found no appreciable $H\alpha$ emission from this galaxy.
However, our SDIG images show several diffuse emission regions
marked by the circles in Fig. 1.

{\bf PGC 51659.} An elongated low surface
brightness galaxy with a bright star projected onto its southeastern
edge. This galaxy is located on the far outskirts of the Cen A group.
It is among the ten richest galaxies of the Local Volume in HI abundance
per unit luminosity. It exhibits appreciable $H\alpha$ emission only on the
northwestern side.

{\bf E 222-010.} A comet-shaped galaxy at low Galactic
latitude $(10^{\circ}$). Its distance was estimated from the line-of-sight
velocity at the Hubble constant $H_0$ = 72 km s$^{-1}$ Mpc$^{-1}$ . The
brightest emission knots in this galaxy are located at its southern boundary.

{\bf E 272-025.} The central region of the galaxy is surrounded by an
extended diffuse halo. Its distance was estimated from the
line-of-sight velocity. Compact emission knots with filaments
are seen only in the central part of the galaxy.

{\bf E~137-018.} An
isolated Sm-type galaxy at low Calactic latitude ($-7.4^{\circ}$). A bright
star is projected near the center. This star produces
a ring-shaped reflex in the telescope's optical system.
The distance to this galaxy, 6.34 Mpc, was first determined by
Karachentsev et al. (2006b).

{\bf IC 4662.} This bright galaxy with extended sites of star formation
is one of the nearest representatives of blue compact galaxies (BCGs).
Karachentsev et al. (2006a) resolved it into stars and determined its
distance, 2.44 Mpc, from RGB stars. Apart from bright compact emission
regions, this galaxy exhibits a network of sinuous $H\alpha$
filaments,
which could be produced by shocks from supernova explosions. The star
formation rate, $9.8\cdot10^{-2}M_{\odot}$yr$^{-1}$, estimated by
Hunter and Elmegreen (2004) agrees well with our data. Hidalgo-Games et al.
(2001) pointed out that the compact double object 80 southeast of the center
of IC~4662 differs greatly from the main galaxy in oxygen abundance.
Being a likely dwarf companion, this object can trigger a starburst in the
main body of IC~4662, as is commonly observed in BCGs (Pustilnik et al.
2001).

{\bf Sag DIG = E~594-04.} This dwarf irregular galaxy in Sagittarius
is the nearest object in our sample (D = 1.04 Mpc). It is located immediately
outside the zero-velocity sphere with a radius of 0.96 Mpc (Karachentsev
and Kashibadze 2006). The tidal index for this relatively isolated galaxy
is --0.3. A single compact emission region lies at the southern
edge of Sag~DIG. According to Hunter and Elmegreen (2004), its
$H\alpha$ flux corresponds to $SFR=1.6\cdot10^{-4}M_{\odot}$yr$^{-1}$,
a value very close to our estimate.

{\bf IC 5052 = E~074-15.} A bright,
isolated edge-on Sd galaxy with many compact HII regions seen in
its disk. Seth et al. (2005) recently estimated the distance to
IC 5052 from RGB stars to be 6.03 Mpc. According to Rossa and
Dettmar (2003), the star formation rate for this galaxy inferred
from its farinfrared flux is 0.08$M_{\odot}$yr$^{-1}$ , in good
agreement with our estimate.

{\bf IC 5152 = E~237-27.} A bright
irregular galaxy at a distance of 2.07 Mpc with many complexes of
blue stars and dust spots. Talent (1980) and Webster
and Smith (1983) performed spectroscopic studies of IC 5152. The
presence of bright field stars makes it difficult to
analyze the structure of this galaxy in detail.

{\bf UGCA 438.} This
isolated irregular galaxy is located on the near outskirts of the
sparse group in Sculptor. A bright star is projected onto its southern
side and produces a ring-shaped reflex in the $H\alpha$  image. Miller
(1996) obtained an $H\alpha$ image of
UGCA 438, but detected no emission. Our $H\alpha$ image exhibits a single
emission region at the southern boundary of the galaxy, the flux
from which is given in Table 2. This galaxy was investigated in the HI
line with the GMRT Indian radio telescope and its HI distribution was
found to be highly asymmetric (Begum 2006).

{\bf E~149-003.} An isolated
dwarf galaxy with an oval central region and an extended envelope (a disk?)
of
low surface brightness. The distance to the galaxy was estimated from
its line-of-sight velocity. In the $H\alpha$ line, the central part of the
galaxy looks like a comet with a bright compact knot on the northwestern
side.

\section{DISCUSSION} As we see from the data presented here, the $H\alpha$
images for the 11 galaxies under discussion differ
greatly in structure, although all of them belong to late morphological
types and are isolated objects. Their emission-line fluxes and star
formation rates also differ by hundreds of times. We used the observed
 star formation rates in galaxies with distances $D < 10$ Mpc from the
literature (van Zee 2000; Gil de Paz et al. 2003; James et al. 2004;
Helmboldt et al. 2004; Hunter and Elmegreen 2004; Karachentsev et al.
2005; Kaisin and Karachentsev 2006). Figure 2
shows the distributions of 154 such galaxies with morphological types
$T > 0$ in absolute magnitude and star formation rate, $SFR$. As follows
from these data, $SFR$ correlates with the galaxy luminosity, so the star
formation rate per unit luminosity remains approximately the same for both
giant and dwarf galaxies. The 11 galaxies that we observed (marked by the
hatched squares) closely follow the general dependence.

To characterize the activity phase of star formation in the galaxies,
we calculated the following two quantities for them:

$ p_*= \lg\{[SFR]\cdot T_0/L_B\}$ and

$f_*=\lg \{M(HI)/[SFR]\cdot T_0\},$

where $T_0$ is the age of the Universe, which is assumed to be 13.7 Gyr
(Spergel et al. 2003), and $M (HI)$ is the total HI mass in the galaxy.
The first parameter shows what fraction of the galaxy's
observed luminosity it would ``gain'' in the Hubble time
at a typical mass-to-light ratio of $1M_{\odot}/L_{\odot}$ and the
current star formation
rate. The second parameter shows how many Hubble times it would take for
the galaxy to run out of its gas at the current star formation rate. In
fact, both dimensionless parameters, $p_*$ and $f_*$,
indirectly characterize the past and future of the star formation
in the galaxy. The last two columns of Table 2 give the parameters
$p_*$ and $f_*$ for the 11 observed galaxies. The solid
straight line in Fig. 2 indicates the $p_*$ = 0 line. The
dash--dotted lines parallel to it correspond to
$p_* = +1$ and $p_*=-1$. As we see,
the $p_*=0$ line is
close to the regression line for the entire set of spiral and irregular
galaxies in the Local Volume.

The star formation in a galaxy can follow
various scenarios: (a) gradual depletion of the initial reserves of gas
in the regime of autonomous steady ``smoldering''; (b)
semiautonomous evolution without any in flow/outflow of
gas, but in the regime of starbursts triggered by external factors
(tides); (c) nonautonomous evolution with a significant
inflow of gas from the intergalactic medium or with the
sweepout of gas from a dwarf galaxy as it passes through the
periphery of a giant galaxy; (d) evolution driven by frequent
galaxy mergers. Some of these scenarios were considered by
Pustilnik et al. (2004), Tutukov (2006a, 2006b), and other
authors. We assume that the distribution of galaxies in the
$\{p_*, f_*\}$ diagram can suggest what type of
evolution agrees better with the available observational data.
Care should be taken that the sample of galaxies is limited by
a well-defined criterion (fixed distance,
flux, and/or morphological type). For
example, our small sample of isolated galaxies shows an anticorrelation
between the parameters $\{p_*$ and $f_*\}$   with a coefficient
of 0.72 (see Fig. 3). This may indicate
that the starburst scenario in isolated galaxies or in galaxies located
in the relatively poor surroundings of neighboring galaxies predominates.
We emphasize that the measurement errors of log $SFR$ are small ($\pm$0.08)
and cannot be responsible for the spread of galaxies along the indicated
diagonal line by an order of magnitude.

\bigskip

{\bf Aknowledgements                        }

We wish to thank
A.V. Tutukov for helpful advice. This work was supported in part by the
Russian Foundation for Basic Research (project no.
04--02--16115).

{\large \bf REFERENCES}

1. A. Begum, private communication (2006).

2. J. S. Gallagher, D. A. Hunter, and A. V. Tutukov, Astrophys. J. 284, 544 (1984).

3. A. Gil de Paz, B. F. Madore, and O. Pevunova, Astrophys. J., Suppl. Ser. 147, 29 (2003).

4. J. F. Helmboldt, R. A. Walterbos, G. D. Bothun, et al., Astrophys. J. 613, 914 (2004).

5. A. M. Hidalgo-Games, J. Masegosa, and K. Olofsson, Astron. Astrophys. 369, 797 (2001).

6. D. A. Hunter and J. S. Gallagher, Publ. Astron. Soc. Pac. 98, 5 (1986).

7. D. A. Hunter and B. G. Elmegreen, Astron. J. 128, 2170 (2004).

8. P. A. James, N. S. Shane, J. E. Beckman, et al., Astron. Astrophys. 414, 23 (2004).

9. S. S. Kaisin and I. D. Karachentsev, submitted to Astrofizika.

10. I. D. Karachentsev and O. G. Kashibadze, Astrofizika 49, 5 (2006) [Astrophys. 49, 3 (2006)].

11. I. D. Karachentsev, V. E. Karachentseva, W. K. Huchtmeier, and D. I. Makarov, Astron. J. 127, 2031 (2004).

12. I. D. Karachentsev, S. S. Kajsin, Z. Tsvetanov, and H. Ford, Astron. Astrophys. 434, 935 (2005).

13. I. D. Karachentsev, A. Dolphin, R. B. Tully, et al., Astron. J. 131, 1361 (2006a).

14. I. D. Karachentsev, R. B. Tully, A. Dolphin, et al., Astron. J., submitted (2006b).

15. B. W. Miller, Astron. J. 112, 991 (1996).

16. S. A. Pustilnik, A. Y. Kniazev, V. A. Lipovetsky, and A. V. Ugrymov, Astron. Astrophys. 373, 24 (2001).

17. S. A. Pustilnik, A. G. Pramskij, and A. Y. Kniazev, Astron. Astrophys. 425, 51 (2004).

18. J. Rossa and R.-J. Dettmar, Astron. Astrophys. 406, 505 (2003).

19. A. C. Seth, J. J. Dalcanton, and R. C. de Jong, Astron. J. 129, 1331 (2005).

20. D. J. Schlegel, D. P. Finkbeiner, and M. Davis, Astrophys. J. 500, 525 (1998).

21. D. N. Spergel, L. Verde, H. V. Peiris, et al., Astrophys. J., Suppl. Ser. 148, 175 (2003).

22. D. L. Talent, PhD Thesis (Rice Univer., Houston, 1980).

23. A. V. Tutukov, Pis'ma Astron. Zh. (in press).

24. A. V. Tutukov, Astron. Rep. 50, 439 (2006).

25. L. van Zee, Astron. J. 119, 2757 (2000).

26. B. L. Webster and M. G. Smith, Mon. Not. R. Astron. Soc. 204, 743 (1983).

\begin{figure}
\centerline{\psfig{figure=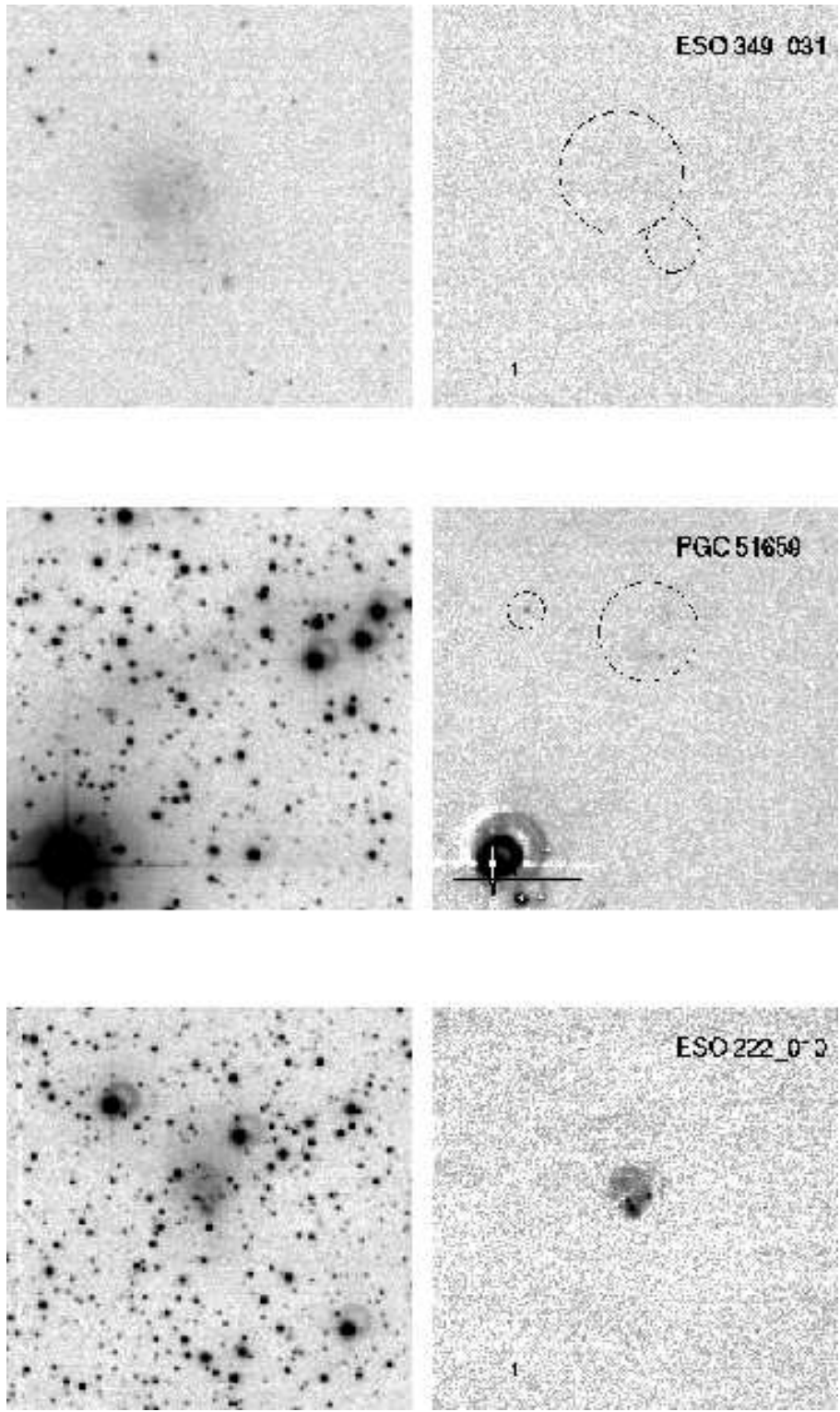,width=12cm}}
\caption{Images of the observed southern galaxies: $H\alpha +$ continuum (left) and $H\alpha -$ continuum (right).}
\end{figure}

\setcounter{figure}{0}
\begin{figure}
\centerline{\psfig{figure=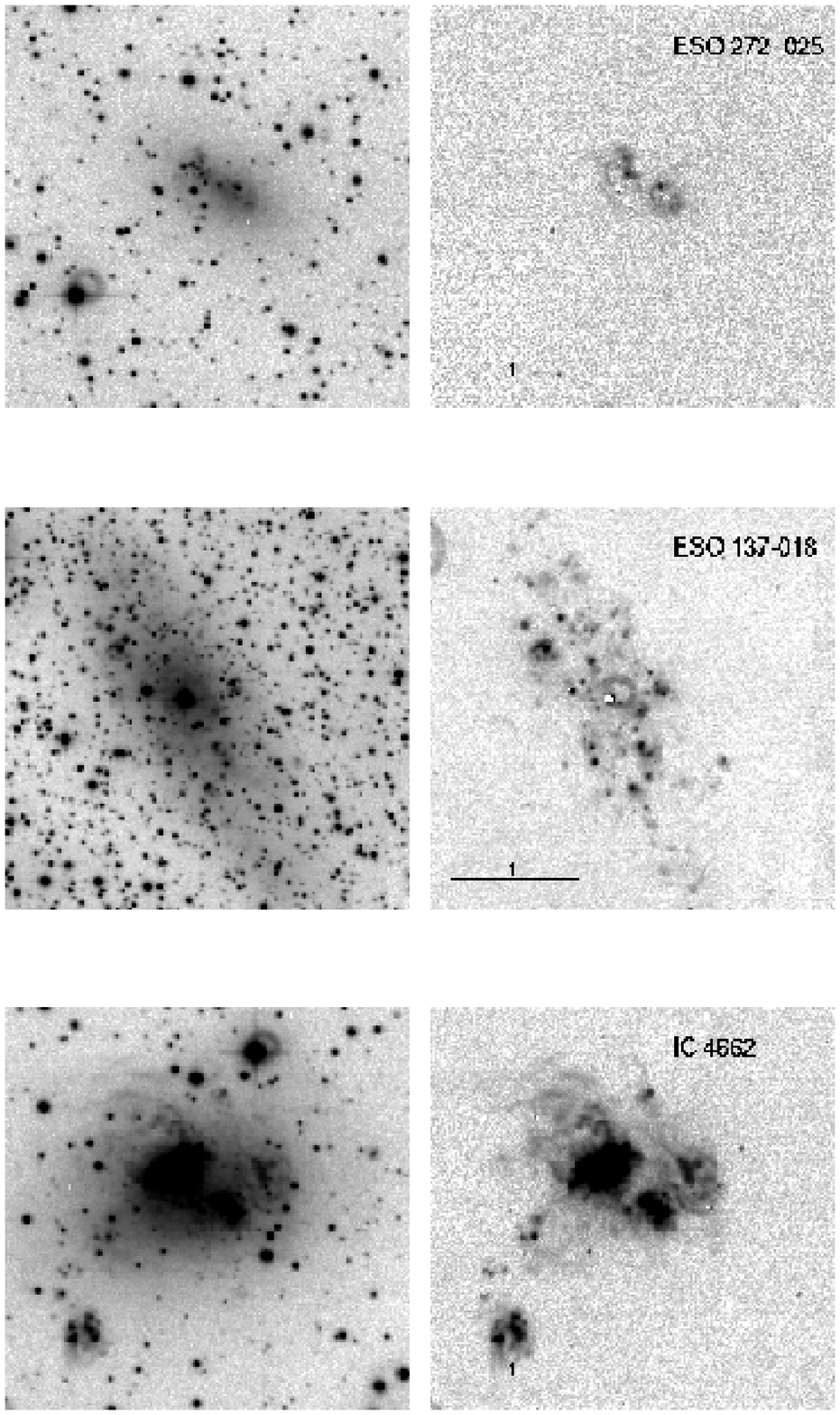,width=12cm}}
\caption{Continued}
\end{figure}

\setcounter{figure}{0}
\begin{figure}
\centerline{\psfig{figure=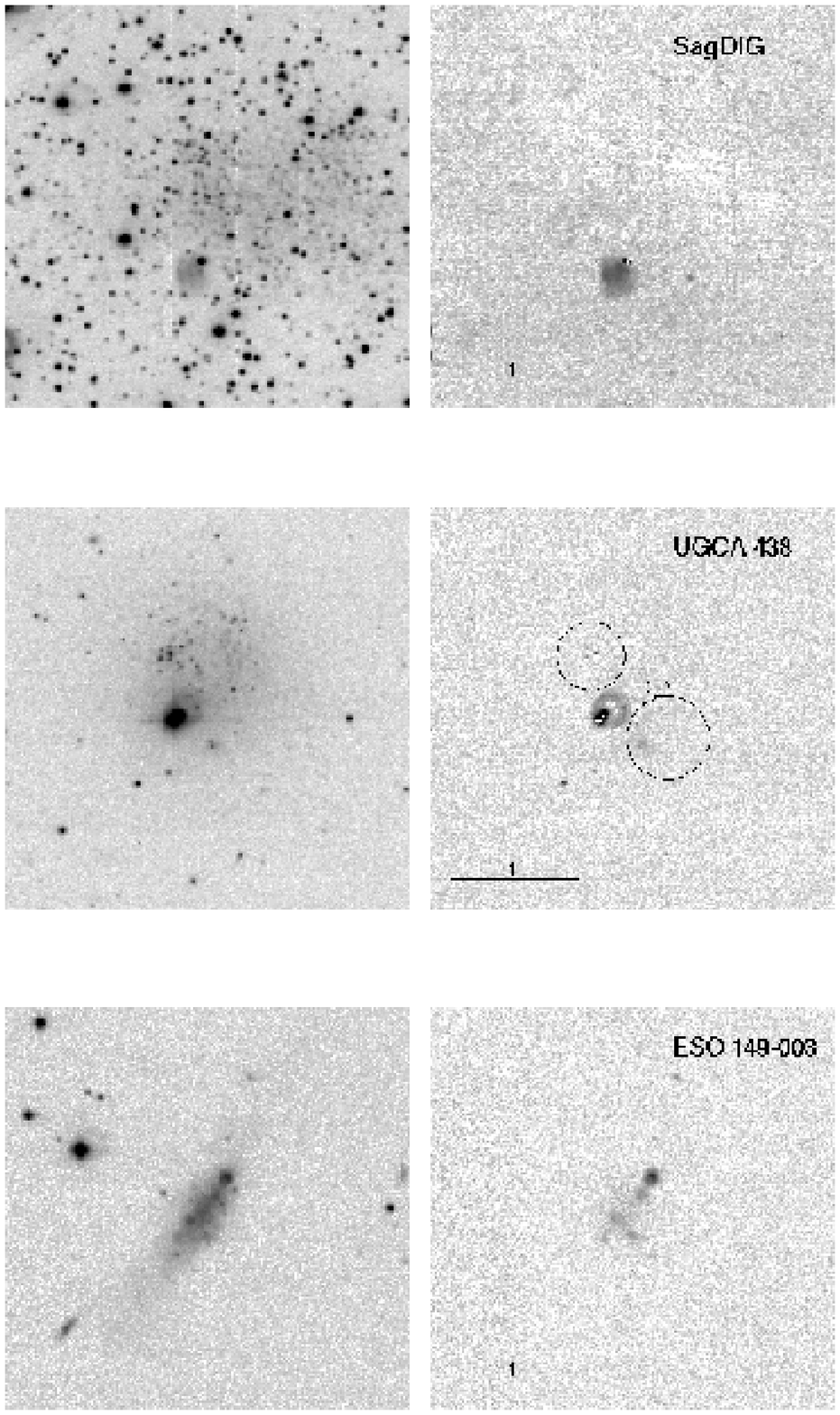,width=12cm}}
\caption{Continued}
\end{figure}
\setcounter{figure}{0}

\begin{figure}
\centerline{\psfig{figure=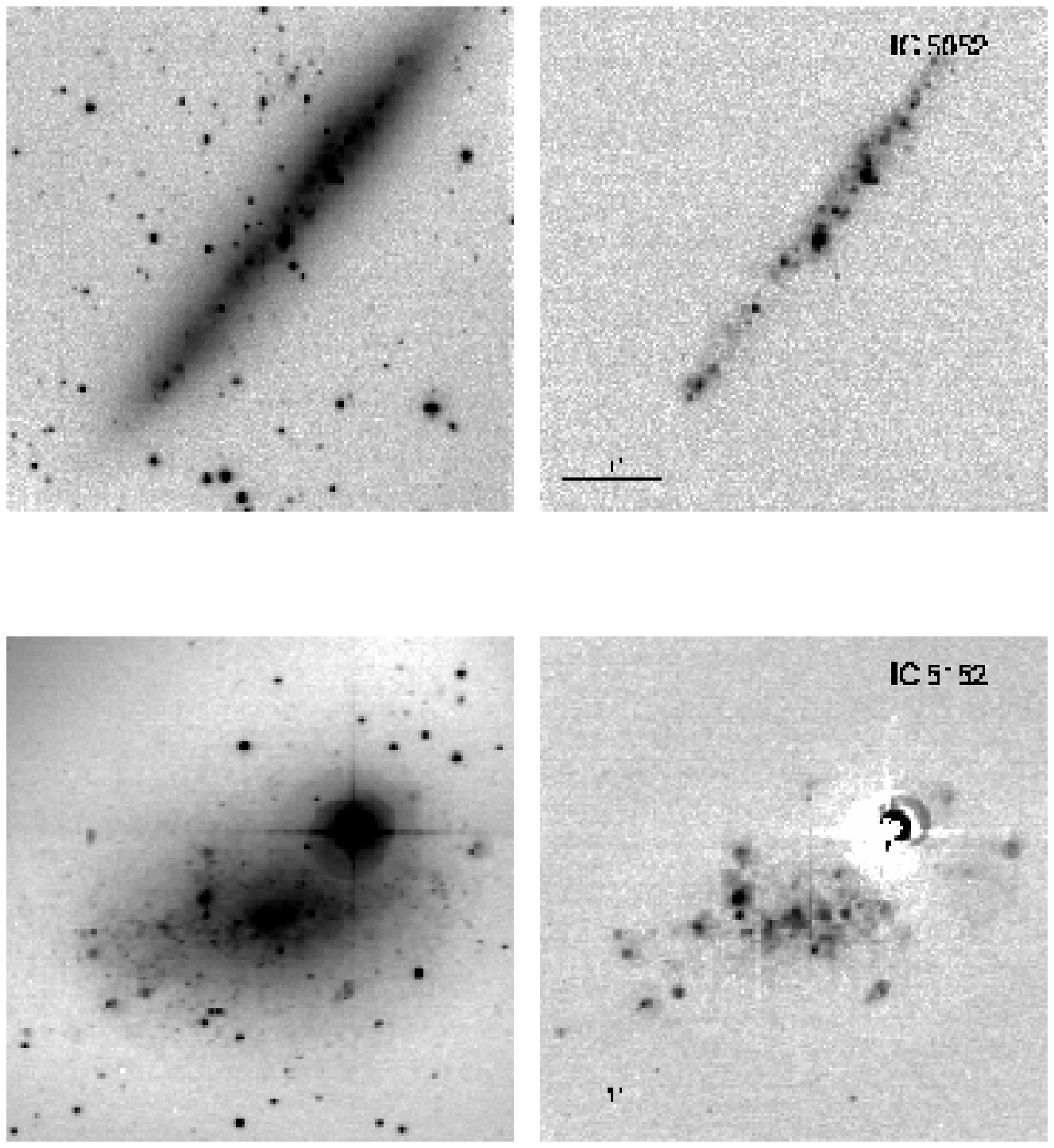,width=12cm}}\
\caption{Continued}
\end{figure}

\begin{figure}
\centerline{\psfig{figure=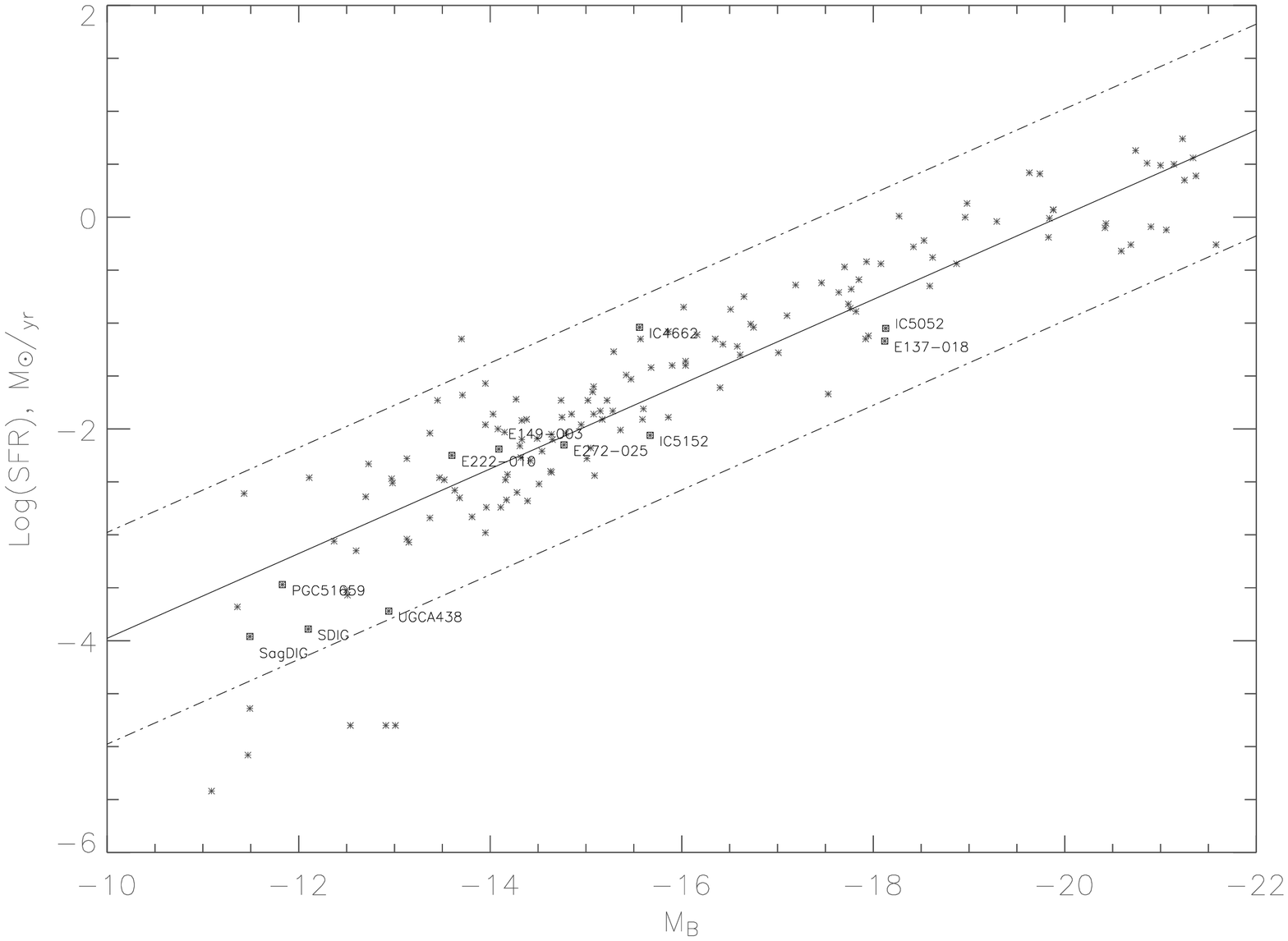,width=18cm,angle=90}}\
\caption{Distribution of the Local Volume galaxies in current star
formation rate and absolute magnitude.}
\end{figure}

\begin{figure}
\centerline{\psfig{figure=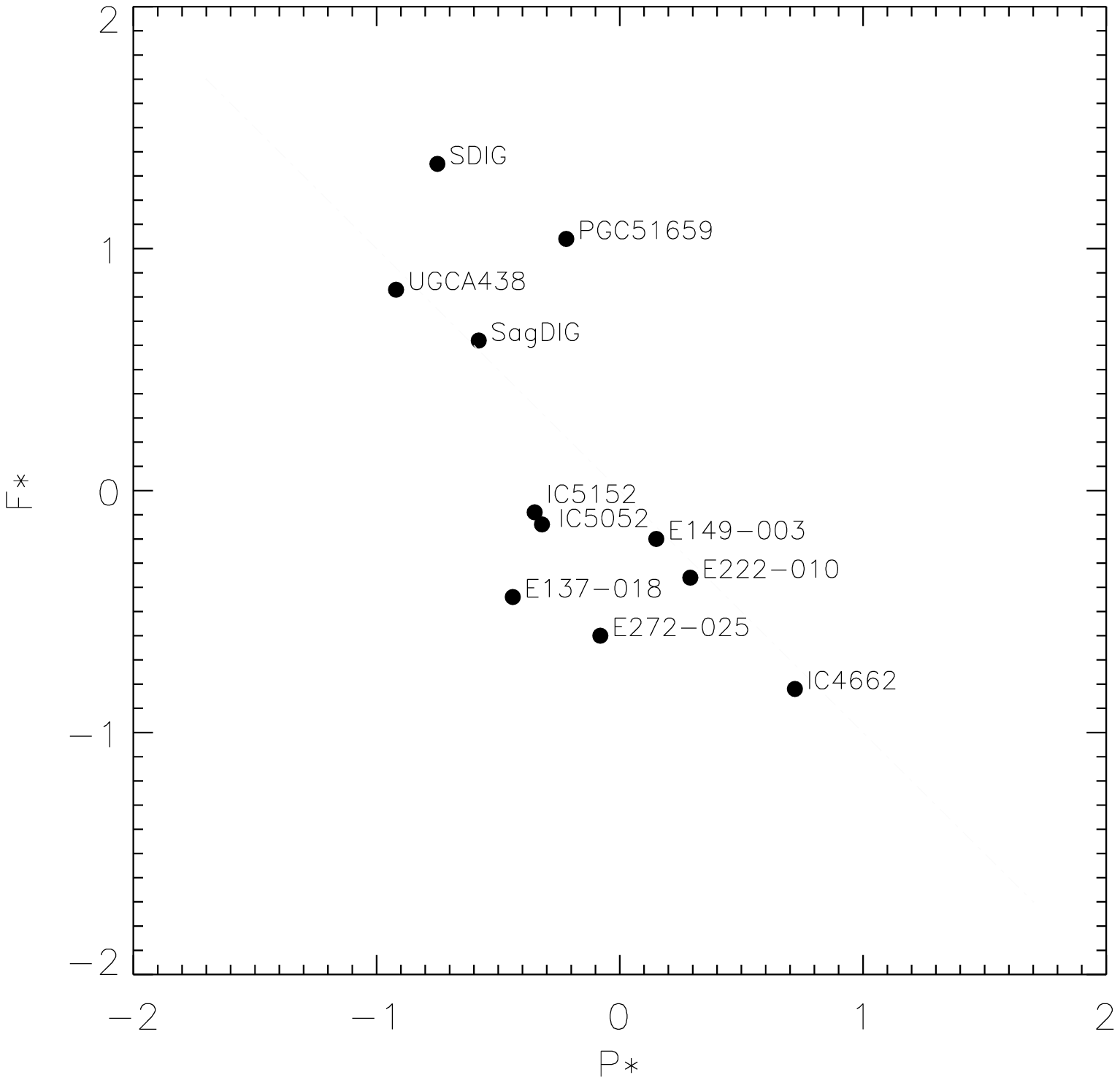,width=18cm}}\
\caption{Distribution of the observed galaxies in $p_*$ and $f_*$}
\end{figure}
\end{document}